1

# TOPICAL REVIEW

# Assembling the puzzle of superconducting elements: A Review

Cristina Buzea [1,2] * and Kevin Robbie [1,2]

[1] Physics Department, Queen's University, Kingston K7L 3N6, Canada
[2] Centre for Manufacturing of Advanced Ceramics & Nanomaterials, Queen's University

**ABSTRACT:** Superconductivity in the simple elements is of both technological relevance and fundamental scientific interest in the investigation of superconductivity phenomena. Recent advances in the instrumentation of physics under pressure have enabled the observation of superconductivity in many elements not previously known to superconduct, and at steadily increasing temperatures. This article offers a review of the state of the art in the superconductivity of elements, highlighting underlying correlations and general trends.

(Some figures in this article are in colour only in the electronic version)

## 1. Introduction

After two decades of focused research on superconductivity in copper oxides or cuprates, interest has recently veered toward simpler materials, such as magnesium diboride and basic elements. Simple materials were recently reported to superconduct at, surprisingly high critical temperatures ($T_c$), 40 K for $MgB_2$ - a binary compound [1], and 20 K for lithium under pressure - the highest $T_c$ for a simple element [2]. During the last few years, the physics community has been delighted with observation of superconductivity in many new elements, some of them under pressure, such as sulphur 17 K [3], oxygen 0.5 K [4], carbon in nanotube 15 K [5] and diamond forms 4K [6], a non-magnetic state of iron 1 K [7], and the light elements lithium 20 K [2] and boron 11 K [8]. While we will soon celebrate a century of superconductivity (Figure 1), these recent discoveries highlight answers to fundamental questions that the elements of the periodic table still conceal. Though thousands of papers have been published on the topic, a general predictive model of superconductivity has eluded us.

__________
* electronic mail: cristi@physics.queensu.ca

Superconductivity is a collective state occurring in the electron population of a material, and its study is an important discipline of condensed matter physics. The occurrence of superconductivity is not restricted to conventional materials, however, but is believed to be a universal phenomenon, arising at all scales from superconducting cosmic strings generated as topological defects in the early universe [9] or superconducting cores in neutron stars [10], down to the fascinating phenomenon of nuclear size superconductivity [11].

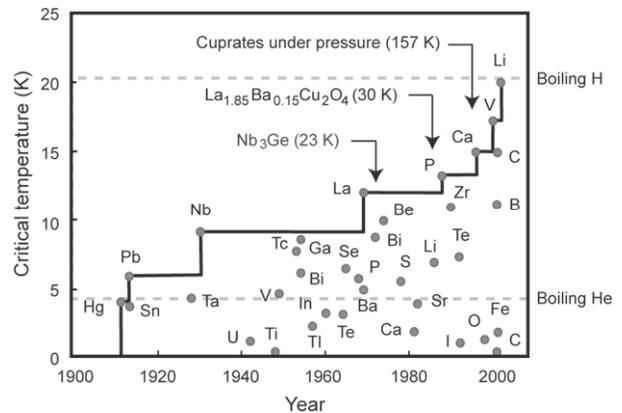

*Figure 1.* Historical development of the critical temperature of simple elements

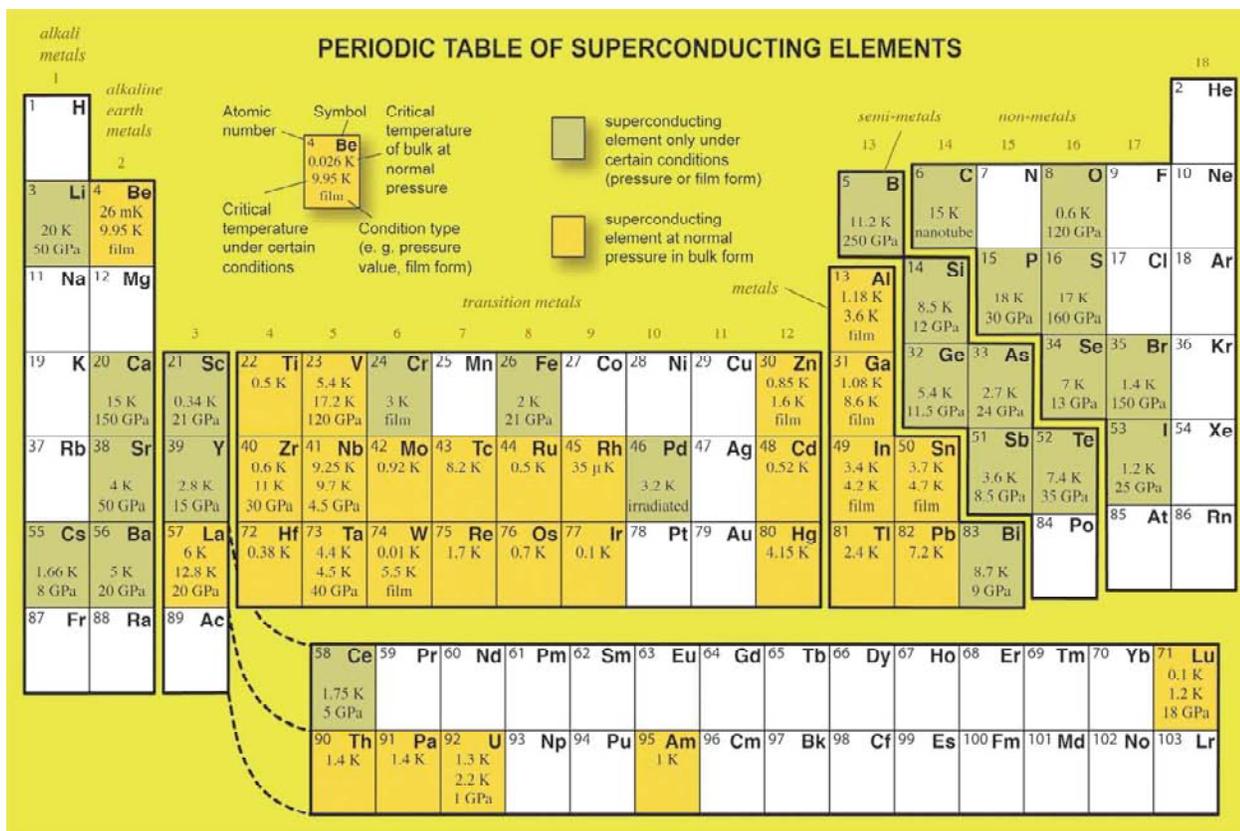

**Figure 2.** *Periodic table of superconducting elements*

Recently experiments and theories on atomic condensation [12] and superconductivity in a two-dimensional electron gas [13] have demonstrated a great variety of correlated states related to superconductivity. First observed in mercury almost a century ago [14], condensed-matter superconductivity remains among the most studied subfields of physics.

This review is organized as a systematic summary of the superconducting behaviour of the chemical elements, referencing the periodic table of Fig. 2. Beginning from hydrogen, ending with iodine, the discussion surveys the table, grouping elements by electronic structure to shed light on the puzzle of superconductivity.

**2. Inducing Superconductivity**

Once thought a rare anomaly, it now appears that superconducting states can be formed in most materials, including thousands of inorganic compounds, most of the natural elements, alloys, oxides, and even organic materials and, significantly, the double-stranded molecules of DNA [15].

If a given element is not superconducting at normal pressure, there are several ways of causing it to superconduct, illustrated schematically in Fig. 3. Transport of electron pairs from a superconductor in close proximity creates a small superconducting interface layer in some materials, a phenomenon called proximity induced superconductivity. Crystal strain or non-equilibrium crystal textures can also induce superconductivity, through applied pressure, quenched growth on a cold substrate, or epitaxial templating in the form of thin films. Chemical doping that affects electron occupancy (charge doping) produces superconductivity in many materials, including diamond. Finally, irradiation induced lattice disorder can also lead to superconductivity through the suppression of spin fluctuations.

The atomic structure of a material determines its superconducting characteristics with superconductivity arising as a collective transport behavior through the electronic potential of the crystal. An illustration of this is the existence of different critical temperature values for polymorphic forms of certain elements, such as lanthanum, which superconducts at $T_c$ = 4.8 K when double hexagonal closed packed, dhcp, compared to $T_c$ = 6 K when face centered cubic, fcc [16].

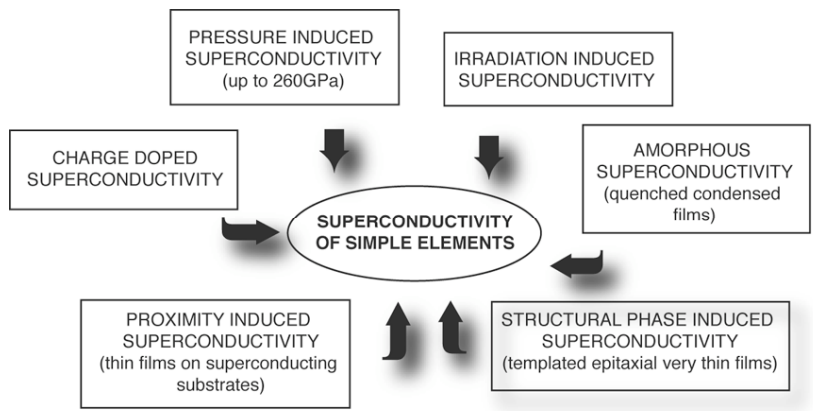

*Figure 3. Techniques used to transform normal elements into superconductors*

## 3. Superconductivity under Pressure

Applying pressure is often the simplest way of modifying the crystalline structure of a material, where the material can undergo structural phase transitions, each structure being characterized by a unique electronic configuration. Phase transformations occur in many elements under compression and some high-pressure phases exhibit superconductivity.

The renaissance in high-pressure research on superconductivity is primarily the result of recent development of high-pressure diamond anvil cells [17]. While the highest attainable pressure in 1968 was approximately 25 GPa employing Bridgman type anvils at liquid helium temperatures [18], today measurements can be performed to a maximum pressure of 260 GPa [8]. This made possible the surprise observation of higher-pressure emergent phases of simple elements that exhibit superconducting properties. Of particular interest is the research of the light elements, as they are predicted to superconduct at record high temperatures [19]. Superconductivity under pressure is seen in many elements, as is summarized in the periodic table of Fig. 2. Some elements transformed to a metallic state without showing signs of superconductivity, e.g. xenon [20], others to semiconductors, e.g. nitrogen [21].

## 4. Superconducting elements

### 4.1. Hydrogen

As the first element in the periodic table, hydrogen attracts much attention, especially due to the predictions of its high-pressure superconductivity at very high temperatures [19]. At moderate pressures and temperatures, H occurs in an insulating molecular phase. Almost seven decades ago hydrogen was predicted to transform under extreme pressure into a monoatomic metallic solid [22], then that metallic H will superconduct in the atomic phase [19], with critical temperatures reaching high values, perhaps to room temperature. The main technological problem encountered in the study of H, the smallest of atoms, is its high diffusivity and chemical reactivity, which make it the most difficult material to contain at high pressure, tending to embrittle and weaken gaskets and anvils. Despite these difficulties, the study of H has generated several interesting results, among which is its transformation into a metallic fluid at 140 GPa [23], and the modification of its optical properties (darkening) at higher pressures [24]. All attempts at demonstrating its superconductivity, however, have had negative results.

### 4.2. Alkali and alkaline metals

Among the alkali metals (Li, Na, K, Rb, Cs, Fr) only Li and Cs were found to superconduct under pressure. Calculations predicted superconductivity of Li at high pressure [25] with a $T_c$ of up to 80 K [26]. Early measurements of Li resistivity under pressure revealed a phase transition accompanied by a sudden drop of resistivity around 7 K, attributed to possible superconductivity [27]. A clear demonstration of Li superconductivity was recently published [2, 28], with $T_c$ increasing with pressure and reaching an unprecedented 20 K for a simple element, under 50 GPa [2]. A plot of its critical temperature dependence on pressure is shown in Fig. 4 together with the behavior of superconducting alkaline earth metals. The experimental behavior of Li is not adequately resolved, and agreement has not been reached on its $T_c$ dependence on pressure [2, 28, 29]. Data measured under almost hydrostatic pressures [29] (compressed in an anvil while surrounded by helium to provide uniform pressure without shear) shows an increase then decrease of $T_c$ with pressure, evidence that superconductivity competes with symmetry breaking structural phase transitions.

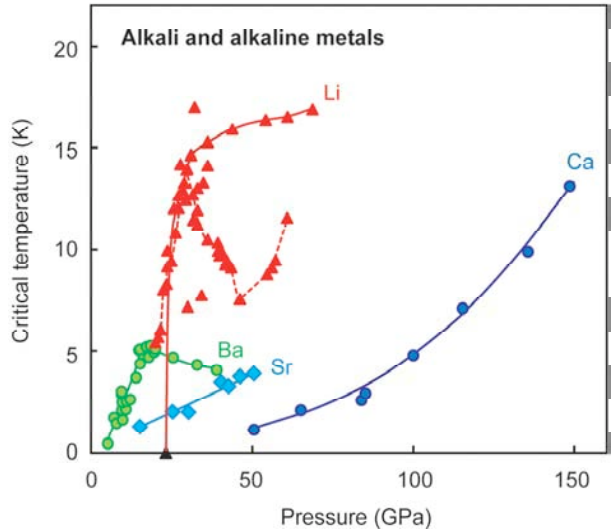

**Figure 4.** Critical temperature dependence on pressure for alkali and alkaline metals, Li [2, 28, 29], Ca [36], Ba [37-43], Sr [37].

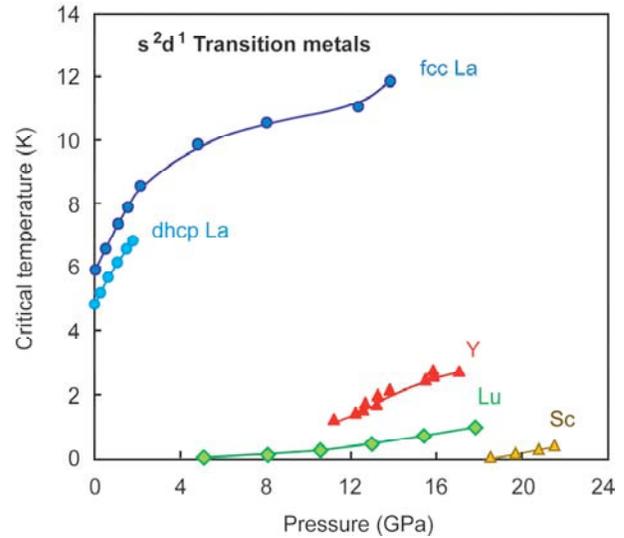

**Figure 5.** Critical temperature dependence on pressure for $s^2d^1$ transition metals, La [16], Lu [44], Sc [44], Y [30].

Cesium under pressure (not shown in Fig. 4) has a much lower critical temperature than Li, with $T_c$ decreasing monotonically from 1.66 K for increasing pressure [30]. For the other alkalis, preliminary calculations suggest that Na might become superconducting under pressure [31], but has yet to be experimentally observed.

Among alkaline earth metals (Be, Mg, Ca, Sr, Ba, Ra) several superconduct: Be, Ca, Sr, and Ba. Beryllium, has a modest $T_c$ of 26 mK in bulk form [32], but attains a $T_c$ of 9.95 K when deposited as a quenched-condensed film [33]. The heavier alkaline earth metals, Ca, Ba, and Sr transform into superconductors only under pressure, explained as an s-d electron transition arising from ion core orthogonality requirements [34]. Calcium [35, 36] and strontium [37] have a positive $dT_c/dp$ curve, while the critical temperature of Ba increases and then decreases [37-42], as seen in Fig. 4. From electronic band structure calculations for Ba [43] it was concluded that the 5d component in the conduction-electron wave function increases strongly under pressure, transforming Ba into a 5d transition metal through s-d hybridization at the Fermi level.

### 4.3. $s^2d^1$ transition metals

Yttrium [30] and its related $s^2d^1$ transition metals Sc, La, and Lu [44], have a positive $dT_c/dp$ slope, as shown in Fig. 5. Lanthanum has two superconducting phases at normal pressure, dhcp at 4.8 K and fcc at 6 K [16].

### 4.4. Transition metals

Most of the transition metals [27 metals from the area cornered by Ti, Hf, Hg, and Zn] are superconducting at normal pressure, the highest critical temperatures being for elements in the fifth group: Nb at 9.25 K [45], V at 5.4 K [46], and Ta at 4.4 K [47]. Niobium also holds the record for the highest critical temperature of an element at normal pressure. No structural phase transitions are observed in Nb and Ta, so changes in $T_c$ are thought to be the result of changes in electronic structure alone, creating a $T_c(p)$ that is nearly constant in Nb from 10 Gpa to 70 GPa, and from ambient to 45 GPa in Ta [45], as seen in Fig. 6. Unlike niobium and tantalum, vanadium has a large positive pressure coefficient of the critical temperature [48], its critical temperature under pressure reaching 17.2 K at 120 GPa, among the highest for simple elements [49]. The increase of $T_c$ is thought to be due to suppression of electron spin fluctuations through broadening of the d-band width [50].

A material less studied since its superconductivity discovery [51, 52], is zirconium, with two superconducting phases, hcp and bcc (body center cubic), reaching a critical temperature of 11 K at 30 GPa in its bcc form [53].

Chromium superconducts at 3 K in an fcc phase fabricated by epitaxially sandwiching Cr between gold layers [54].

Palladium is a metal with particularly interesting properties. From its high electronic density of states at the Fermi surface and its phonon spectrum, one would expect strong electron-phonon coupling, and therefore a reasonable high superconducting transition temperature. Experimentally, however, superconductivity does not occur down to 1.7 mK [55] due to spin fluctuations, though palladium can be transformed into a superconductor with a $T_c$ of 3.2 K through the introduction of lattice disorder via low-temperature irradiation with $He^+$ ions [56], in qualitative agreement with theoretical predictions for superconductivity in palladium without spin fluctuations [56].

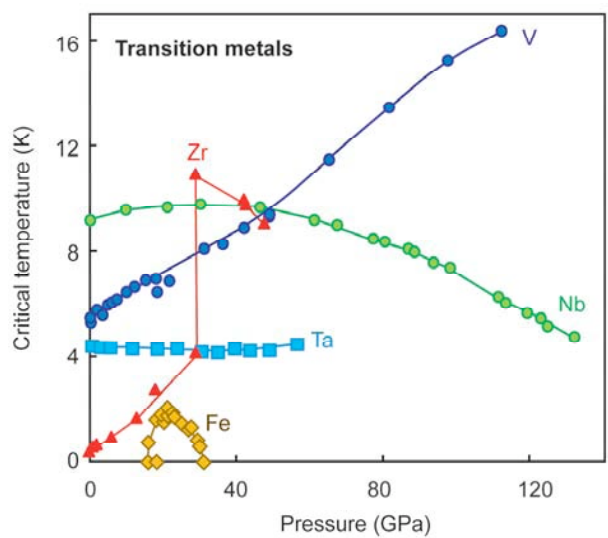

**Figure 6.** Critical temperature dependence on pressure for transition metals, Zr [52, 53], V [48-50], Nb [45], Ta [45, 67], and Fe [7].

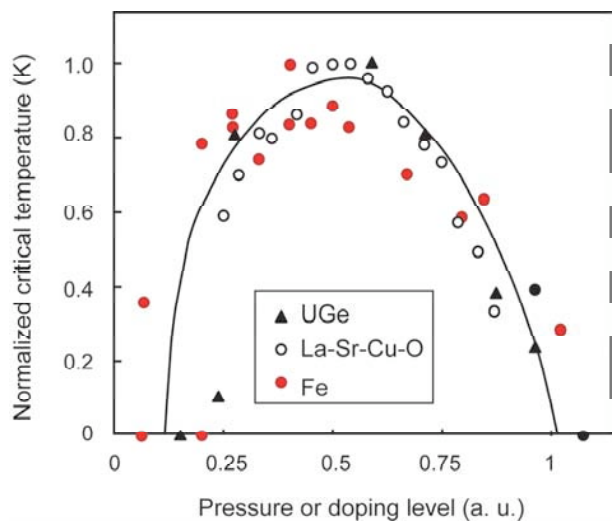

*Figure 7. Normalized critical temperature variation with pressure and doping level for materials with ferromagnetic or antiferromagnetic behavior. Data are taken from references [59], [58] and [7] for UGe2, La-Sr-Cu-O, and Fe, respectively.*

*4.5. Superconductivity and magnetism*

The correlation between superconductivity and magnetism makes the study of ferromagnetic metals (Fe, Co, Ni, Gd) particularly important. Iron is the classic magnetic metal, being strongly ferromagnetic at standard conditions in its bcc phase. In this ferromagnetic form iron does not superconduct to the lowest temperatures attainable. At pressures above 10 GPa iron assumes an hcp structure, believed to be non-magnetic. The superconductivity of the iron hcp phase was predicted [57] and was recently confirmed experimentally [7]. As illustrated in Fig. 6, the $T_c(p)$ diagram for iron has a specific bell-shaped curve. Interestingly, this phase diagram for iron [7] shows a striking resemblance to the critical temperature versus doping or carrier concentration of superconductors with antiferromagnetic parent compounds - $La_{2-x}Sr_xCu_2O_4$ [58], or $UGe_2$ [59], as illustrated in Fig. 7. While their relationship is not fully understood, it was believed that ferromagnetism and superconductivity were competing mechanisms, and the superconductivity of iron came as a surprise. Perhaps the explanation lies in theoretical model of magnetism mediated superconductivity [59]. The validity of this model was questioned until recently, when superconductivity was observed in $UGe_2$, on the border of ferromagnetism within the same electrons that produce band magnetism [59]. These recent discoveries of superconductivity in iron [7], as well as in cobalt compounds [60, 61], re-focuses attention on magnetic mediated superconductivity as an alternative to phonon mediated.

*4.6. Metals*

Metals from groups 13-15 (Al, Ga, In, Tl, Sn, Pb, Bi) situated on the right edge of the transition metals in the periodic table show several common characteristics. First, they all superconduct at normal pressure, except bismuth, which is almost a semi-metal. Second, the $T_c$s of the elements decrease with increasing pressure, as shown in Fig. 8 for Al [62], Sn [63], Pb [40, 64-67] and Bi [39, 40], the derivative $dT_c/dp$ having similar values. In addition, the critical temperature seems to increase from group 13 to 15.

Aluminum superconducts below 1.18 K [62], and when in mesoscopic wire form shows a peculiar resistance peak in the superconducting state near $T_c$, attributed to thermal fluctuations producing phase slips of the superconducting order parameter [68].

Gallium has a critical temperature of 1.08 K in bulk form, and in amorphous film form, quench-condensed on liquid-helium cooled substrates, attains 8.6 K [69].

Bismuth is an example of those elements that do not display superconductivity under ordinary circumstances, but undergo superconducting phase transitions in a fcc phase [70] after being subjected to hydrostatic pressure. Bi superconducts with a $T_c$ of 8.7 K at 9 GPa [40], with a $T_c$ = 4 K [70] in thin film form on Ni substrates, and with $T_c$ = 6.2 K [71] when amorphous.

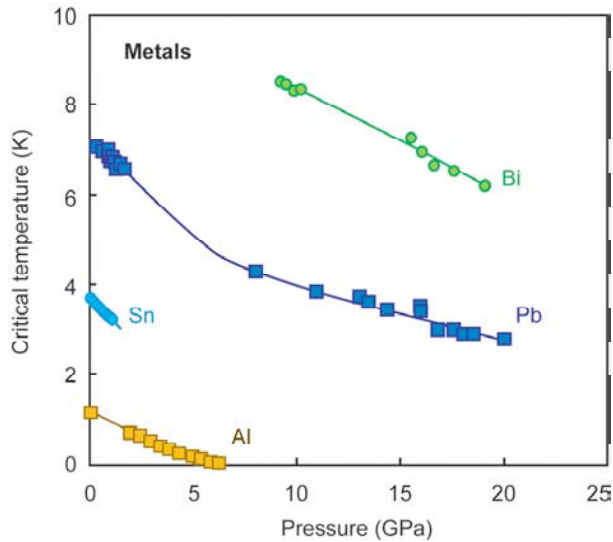

*Figure 8. Critical temperature dependence on pressure for metals, Al [62], Sn [63], Pb [40, 64-67], Bi [39, 40].*

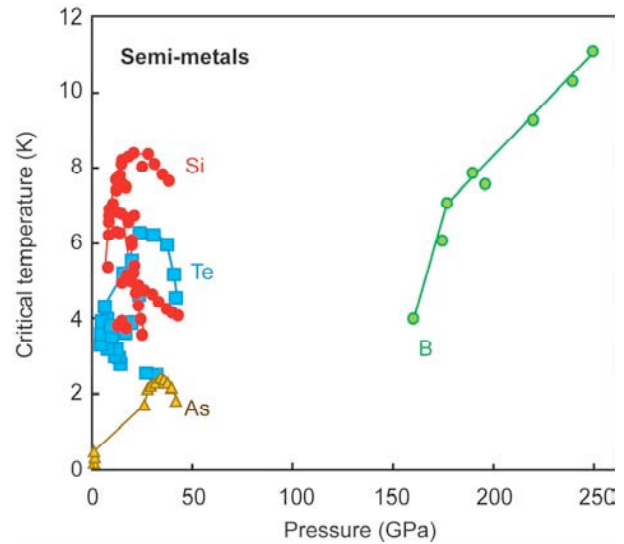

**Figure 9.** Critical temperature dependence on pressure for semi-metals, B [8], Si [72-74], As [75, 76], Te [77-79].

*4.7. Semi-metals*

A general characteristic of semi-metals is that they do not show superconductivity at normal pressure, but superconduct when compressed. Most of them have structural phase transitions within the studied pressure range, and exhibit different $T_c$s for different phases. As a common feature of Si [72-74], As [75, 76], and Te [77-79], the critical temperature increases with pressure, followed by saturation and a decline (Fig. 9). The recent research of superconductivity in the lightest semi-metal - boron, probably triggered by the announcement of $MgB_2$ superconductivity at nearly 40 K [1], spurred interest in superconducting physics on boron-related compounds [80]. Boron superconductivity was observed with a maximum $T_c$ of 11.2 K at 250 GPa [8], as shown in Fig. 9.

*4.8. Non-metals*

Among non-metals, carbon (in the form of diamond and nanotubes), oxygen, bromine, and iodine have recently been transformed into superconductors, while the superconductivity under pressure of phosphorus, sulfur, and selenium was observed much earlier (Fig. 10). Nitrogen was transformed into a semiconductor under pressure without exhibiting superconductivity [21].

Carbon likely occurs in the widest variety of elemental forms, the most known allotropes being graphite, diamond, C60 carbon spheres (Buckminster-fullerenes), and carbon nanotubes derived from curved graphene sheets. The possibility of superconductivity in carbon nanotubes was predicted in 1995 with $T_c$ expected to be inversely proportional to the nanowire diameter, an effect related to electron-phonon interactions [81]. In 2001 Kociak et al. [82] reported superconductivity below 0.55 K in ropes of single-walled carbon nanotubes with low-resistance contacts to non-superconducting metallic pads. Later the same year Tang et al. [5] succeeded in observing superconductivity in one-dimensional 0.4 nm diameter single-walled carbon nanotubes encapsulated in channels of zeolite crystals with a record high $T_c$ of 15 K. The latest surprise from superconductivity in carbon-based materials came from boron doped diamond [6]. The discovery of superconductivity in a diamond-type structure suggests that Si and Ge, also having a diamond crystal structure, might superconduct under appropriate doping conditions.

Since the discovery of superconductivity in phosphorous [83] at 5.8 K under 17 GPa its critical temperature was increased under 30 GPa pressure to 18 K [84]. The superconducting transition temperature of phosphorus under pressure depends on the path in the pressure-temperature diagram [84-86]. Black insulating phosphorous with an orthorhombic structure is the most stable form at ambient conditions. When pressure is increased at liquid-helium temperature black phosphorus transforms to a metallic phase with a simple cubic structure with a $T_c$ up to 10.5 K [85]. If red phosphorus is used as the starting material and pressurized at 4.2 K, the $T_c$ increases up to 13 K with the onset of transition at 18 K under 30

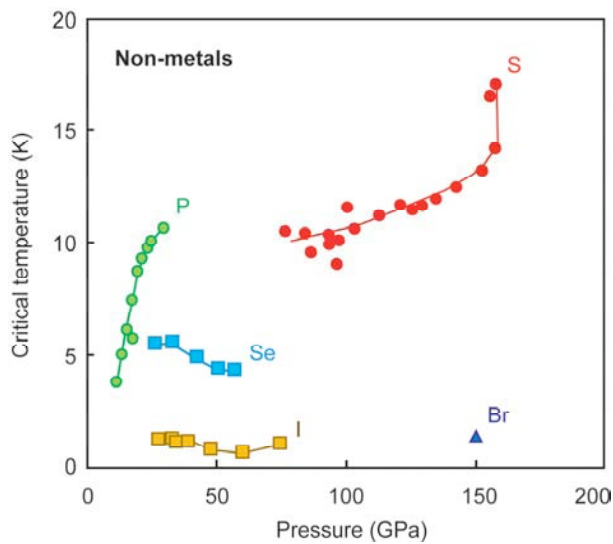

**Figure 10.** Critical temperature dependence on pressure for non-metals, P [86], S [3, 93], Se [77], Br [93], I [92].

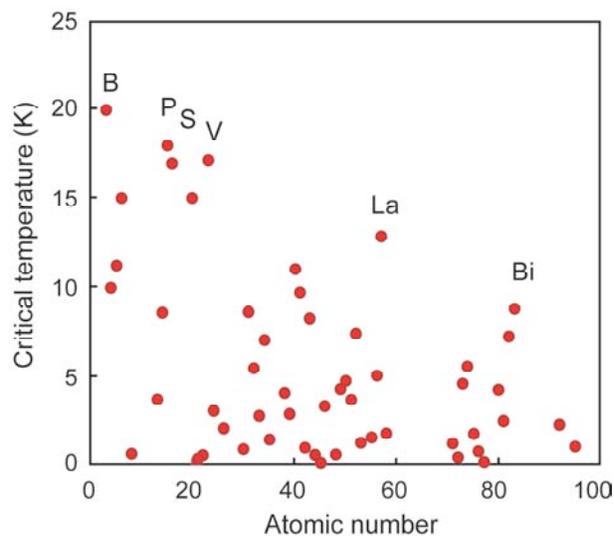

*Figure 11. Highest critical temperature of simple elements as a function of the atomic number.*

Gpa. The resistance transition versus temperature exhibits a drop with a finite temperature width, with onset at 18 K and zero resistance at 8 K. A new simple bcc structure has been observed above 262 GPa [87], giving hope that a similar or higher $T_c$ could be achieved in this phase. It was suggested that this bcc phase of phosphorous might be stabilized at ambient conditions using suitable templates such as V(100), Fe(100), or Cr(100) substrates [88]. This could lead to a breakthrough in technological spintronic applications, where the combined spin and superconducting degrees of freedom will provide a new level of functionality for microelectronic devices.

The first studies of superconductivity of sulfur were performed in 1978 with reports of transitions into the superconducting state at 5.7 K [89] and 9.8 K [90]. Recent studies of compressed sulfur [3, 91] show that the element transforms to a superconductor with $T_c$ increasing with pressure to 17 K at 160 Gpa (Fig. 10). This is the third highest reported transition temperature for an elemental solid. This behavior contrasts with the negative $dT_c/dp$ observed at much lower pressures in the heavier superconducting chalcogens Se and Te [77].

The successful searches for superconductivity in the halogen group were iodine [92] and bromine [93]. In the case of bromine under pressure, a molecular to monoatomic phase transition takes place at 80 GPa. At pressures higher than 90 GPa bromine becomes a superconductor [93].

Oxygen is known to show magnetic properties at low temperatures, so was not expected to superconduct. At pressures exceeding 95 GPa, solid molecular oxygen undergoes a phase transition and becomes metallic. At around 100 GPa solid oxygen becomes superconducting below 0.6 K, as revealed by resistive and magnetic measurements [4]. It seems that the oxygen superconductivity is present even in the molecular metallic state, in contrast with iodine and bromine. The transition temperature of oxygen is rather low compared to other elements in the 16th series: S, Se, and Te which superconduct at 17, 7, and 7.4 K, respectively, in spite of all expectations from both the structural sequence and the increase of $T_c$ for this group.

None of the noble gases (He, Ne, Ar, Kr, Xe, Rn) are known to superconduct. A report on xenon indicates its transforms into a metallic state but without showing signs of superconductivity [20].

## 5. Critical temperature correlations

As a general trend, the maximum critical temperature of simple elements scales with the atomic number Z, the highest $T_c$'s being achieved for low Z (Fig. 11), implying that light elements are the best candidates for the utmost critical temperatures. Other correlations of $T_c$ with normal state properties [97], such as bulk modulus, work function, Hall coefficient, Debye temperature, etc. should be re-evaluated in the light of the most recent findings in the critical temperature of simple elements. An interesting path would be a search for possible correlation of $T_c$ with Hall coefficient changes under pressure.

## 6. Conclusions

As future directions, further theoretical and experimental studies of superconductivity at pressures exceeding 50 GPa for most of the elements are needed.

While applying pressure to a material is an established way of exploring new crystalline structures, recent thin film fabrication methods have shown great utility in growing unique and non-equilibrium crystals through sandwiched epitaxy [54] or control of geometrical effects in vapor aggregation [94, 95]. These methods should be further employed in the search for new superconducting phases.

The study of elemental superconductors has an enormous impact on our understanding of superconductivity. Current work is laying the foundations of elemental superconductor nanowires applications and further elucidation of the low dimensionality effect on their properties. Advances in materials and nanomaterials fabrication and characterization will certainly lead to observation of superconductivity in elements at higher temperatures, creating new opportunities for applications such as spintronics [88] or DNA integrated molecular electronics [96]. As suggested by the fast pace of discoveries in this field of superconductivity, it seems the periodic table of superconducting elements will likely need updating in the near future.

**Acknowledgements**

*We thank I. Pacheco for stimulating discussions and invaluable suggestions, and J. E. Hirsch, J. M. Fraser, M. J. Stott, and K. Shimizu for helpful comments on the text. This research was supported in part by the Natural Sciences and Engineering Research Council of Canada (NSERC).*